\documentclass[pre,preprint,showpacs]{revtex4-1}
\usepackage[dvipdfmx]{graphicx}
\usepackage{amsmath,amssymb}
\usepackage{dcolumn}% Align table columns on decimal point
\usepackage{bm}% bold math
\usepackage{hyperref}% add hypertext capabilities

\begin{document}
\title{Scaling analysis of stationary probability distributions of random 
walks on one-dimensional lattices with aperiodic disorder}

\author{Hiroshi Miki}
\altaffiliation{Present address: Research Institute for Humanity and Nature, 
457-4 Motoyama Kamigamo, Kita-ku, Kyoto 603-8047, Japan.}

\affiliation{
Department of Applied Science for Electronics and Materials,\\ 
Interdisciplinary Graduate School of Engineering Sciences, \\
Kyushu University, 6-1 Kasuga-Koen, Fukuoka 816-8580, Japan
}

\date{\today}

\begin{abstract}
Stationary probability distributions of one-dimensional random walks on 
lattices with aperiodic disorder are investigated. 
The pattern of the distribution is closely related to the diffusional 
behavior, which depends on the wandering exponent $\Omega$ of the background 
aperiodic sequence: If $\Omega<0$, the diffusion is normal and the 
distribution is extended. If $\Omega>0$, the diffusion is ultraslow and the 
distribution is localized. If $\Omega=0$, the diffusion is anomalous and the 
distribution is singular, which shows its complex and hierarchical structure.  
Multifractal analysis are performed in order to characterize these 
distributions. Extended, localized, and singular distributions are clearly 
distinguished only by the finite-size scaling behavior of $\alpha_{\rm min}$ 
and $f(\alpha_{\rm min})$. The multifractal spectrum of the singular 
distribution agrees well with that of a simple partitioning process.
\end{abstract}

\maketitle

\section{Introduction}
Random walks have long been one of the most fundamental processes both in physics 
and mathematics\cite{BG}. 
Due to their simplicity and interesting and rich structure, they 
provide the basis for understanding many kinds of physical phenomena, including  
transport processes, fluctuating time series, relaxation processes, and pattern 
formation. It is well known that for a random walk on a symmetric 
and homogeneous background, the average displacement vanishes and the average 
mean-square displacement scales linearly with time:
\begin{eqnarray}
<X(t)>=0,
\\
<X^2(t)> \sim t.
\label{normaldif}
\end{eqnarray}

On the other hand, it is also known that the diffusional behavior of the system 
is strongly and qualitatively modified by disorder, especially when the spatial 
dimension is low. In a one-dimensional random walk with random disorder, the 
diffusion is strongly suppressed and the average mean-square displacement grows 
on a log-time scale
\cite{Sinai, Derrida}:
\begin{equation}
<X^2(t)> \sim (\log t)^4,
\label{logdif}
\end{equation}
which is called ultraslow diffusion.   

In this article we consider systems with aperiodic disorder. An aperiodic 
disorder is generated by a certain deterministic rule. 
It is this point that distinguishes aperiodic from random disorder. We expect 
that the behavior of systems with aperiodic disorder is, in general, 
intermediate between that of homogeneous system and that of system with 
random disorder. 
In fact, for a random walk on a certain particular one-dimensional lattice 
with aperiodic disorder, it was reported in ref.\cite{ITR} that an anomalous 
diffusion may occur. This is characterized as 
\begin{equation}
<X^2(t)> \sim t^\phi, \quad \text{with }0<\phi<1.
\label{anomdif}
\end{equation}
Interestingly these normal, ultraslow, and anomalous diffusion are observed 
in dynamical deterministic maps\cite{DK}. 
Recently an unified understanding of the diffusional 
behavior described by Eqs.(\ref{normaldif})-(\ref{anomdif}) has been 
attempted from the point of view of a weakly chaotic regime 
of a deterministic map\cite{RV}. In addition to the theoretical and 
mathematical interest, systems with aperiodic disorder have been fabricated 
artificially\cite{apexp}.

We investigate the structure of the stationary probability 
distribution of a random walk on a one-dimensional lattice with aperiodic 
disorder. For a one-dimensional lattice, aperiodic disorder is expressed by a  
corresponding aperiodic sequence. 
It is the wandering exponent $\Omega$ that characterizes an aperiodic disorder 
and affects the diffusional behavior. It determines how the geometrical 
fluctuation of the sequence $\Delta$ scales with the length of the sequence $L$, 
$\Delta \sim L^{\Omega}$\cite{ITR,Luck,Hermisson}: 
If $\Omega$ is negative, the geometrical fluctuation is bounded and the 
effect of the disorder becomes smaller as the size of the system grows. 
Therefore, the diffusional behavior becomes qualitatively similar to that on 
a homogeneous background. On the other hand,  if $\Omega$ is positive, 
the effect of the disorder becomes stronger with an increase in the system size. 
If $\Omega$ vanishes, the effect of the disorder is almost independent of 
the system size - the fluctuation grows logarithmically. 
In this case we observe anomalous diffusion, which is written as 
Eq.(\ref{anomdif}).

Therefore, we expect that the stationary probability distribution will show 
a characteristic pattern. Furthermore, we expect the pattern to depend 
on only the wandering exponent and to correspond to the diffusional behavior. 
We do not expect it to depend on the details of the aperiodic sequence.

Here we consider random walks on one-dimensional lattices, for which 
the disorder is constructed by the Thue-Morse (TM), the Rudin-Shapiro (RS), 
and the paperfolding (PF) sequences, which are taken as representative examples. 
The TM, RS, and PF sequences have negative, positive and vanishing wandering 
exponents, respectively. They have several properties in common: 
1) They are binary sequences, {\it i.e.}, they are 
composed of two types of symbols, $A$ and $B$; 2) They are constructed 
systematically from the initial sequences and by the substitution rules; 
3) The ratio of the number of $A$ to that of $B$ converges to unity in the 
limit of infinite length.
In these cases the geometrical fluctuation of the sequence is given by  
the difference between the number of $A$ and that of $B$.  

Multifractal analysis will be used to characterize the structure of 
the stationary probability distribution. Suppose that a probability distribution 
is given and the support of the distribution is covered with patches of size 
$\epsilon$. Let $p_j(\epsilon)$ be the measure assigned to  the $j$-th patch.   
It is expected that the measure scales with $\epsilon$ as
\begin{equation}
p_j(\epsilon) \sim \epsilon^{\alpha_j},
\end{equation}
where $\alpha_j$ is the singularity exponent. 
It is also expected that the number of patches 
which takes the value of the singularity exponent between $\alpha$ and 
$\alpha + d \alpha$ also scales as 
\begin{equation}
N(\alpha)d\alpha \sim \epsilon^{-f(\alpha)} d\alpha,
\end{equation}
where $f(\alpha)$ is, roughly speaking, the fractal dimension of the set of 
patches with $\alpha$. Multifractal analysis has been applied to characterize 
the scaling structure of various distributions including those of the quantum 
localization problem\cite{HK}, energy dissipation in turbulence\cite{CJ,CMJS}, 
and the sidebranch structure of dendrites\cite{MH}. Since it is so widely 
applicable, we expect that multifractal analysis will also be a good tool 
for our investigation. 

The rest of this paper is organized as follows: In Section II we present our 
model. First we describe our random walk model on a one-dimensional 
lattice with aperiodic disorder. Next we introduce the aperiodic sequences 
mentioned above and refer to their properties necessary for our study. 
In Section III 
multifractal analysis is performed. We describe the finite-size scaling 
formulation and discuss the criterion for classifying the localization 
property of the distribution and the finite-size effect. After 
briefly discussing the relationship with the inverse partitioning ratio, 
which characterizes the localization property, we present our results and 
discussion. Section IV is dedicated to the summary and future outlook.  

\section{Model}
\subsection{Random walk on a one-dimensional disordered lattice
\label{1drw}}
Consider a one-dimensional random walk with only nearest neighbor hopping 
allowed. The time evolution of the probability for the particle to be 
on site $j$ at time $t$, $p_j(t)$, is described by the master equation:  
\begin{equation}
\frac{\partial p_j(t)}{\partial t}
= w_{j-1,j}p_{j-1}(t) + w_{j+1,j}p_{j+1}(t)
-(w_{j,j-1}+w_{j,j+1})p_j(t),
\label{mastereq}
\end{equation}
where $w_{j,k}$ denotes the transition rate for the particle to hop from 
site $j$ to site $k$. Two transition rates are assigned 
to the $j$-th bond, which connects the $j$-th site to the $(j+1)$-th site. 
One is the forward rate $w_{j,j+1}$, and the other is the backward rate 
$w_{j+1,j}$. These transition rates are 
generally not symmetric, {\it i.e.}, $w_{j,j+1} \ne w_{j+1,j}$. 
Interestingly, in ref.\cite{ITR}, it is pointed out that the 
master equation Eq.(\ref{mastereq}) is equivalent to the transverse-field 
Ising model.  

Let us take a binary sequence $S$, for example, $S=ABAABAABAB \cdots$. 
For this sequence, the transition rates are assigned as
\begin{equation}
\frac{w_{j+1,j}}{w_{j,j+1}}=
\begin{cases}
a, &\quad \text{the $j$-th bond is type $A$},
\\
b, &\quad \text{the $j$-th bond is type $B$}. 
\end{cases}
\label{hoprates}
\end{equation} 
If $a=b$, the one-dimensional lattice is homogeneous. 
It is apparent that the properties of the sequence $S$ strongly 
affect the behavior of the random walk. 

By definition, an aperiodic sequence has infinite length. We consider an 
aperiodic sequence which is constructed systematically by substitution rules 
and replace the fully aperiodic sequence with a finite approximant $S_n$ of 
finite length $L$, where $n$ denotes the generation of the approximant. 
We impose the periodic boundary conditions, $p_{j+L} \equiv p_j$, 
$w_{j+L,j+L+1}=w_{j,j+1}$, and $w_{j+L,j+L-1}=w_{j,j-1}$. 
The aperiodic sequence is recovered in the limit as $L \rightarrow \infty$. 

We now consider the stationary probability distribution, $dp_j/dt=0$. 
For simplicity, we set $w_{j,j+1}=1$ for all $j$. 
The exact expression of the stationary probability is obtained\cite{Derrida} 
from the proportional relation
\begin{equation}
p_j \propto 1+\sum_{k=1}^{L-1}\prod_{l=1}^k w_{j+l+1,j+l},
\label{solution}
\end{equation}
and the normalization condition
\begin{equation}
\sum_{j=1}^{L} p_j=1.
\label{normalization}
\end{equation}
   
From this stationary probability, we can obtain the drift 
velocity (and the diffusion constant if Eq.(\ref{normaldif}) 
holds)\cite{ITR,Derrida}. The drift velocity $v_d$ is given as
\begin{equation}
v_d  \propto 1-\prod_{j=1}^L w_{j+1,j},
\label{driftvel}
\end{equation}
If $v_d$ is sufficiently large, the stationary probability distribution is 
extended throughout the system independently of the properties of the 
disorder. On the other hand, if $v_d$ is small, 
diffusion is dominant and then the disorder of the lattice strongly affects 
the behavior of the system. Since we are interested in how the structure of the 
probability distribution is related to the diffusional behavior, we will 
consider only the case with vanishing drift velocity, {\it i.e.}, 
from Eq.(\ref{driftvel}),
\begin{equation}
\prod_{j=1}^L w_{j+1,j} = 1.
\label{vanishing_vd}
\end{equation}
For a given $a$, $b$ is given as a function of $a$ so that 
Eq.(\ref{vanishing_vd}) hold. 

\subsection{Aperiodic sequences}
In this subsection, we review three aperiodic sequences, each of which we 
consider below for the disorder of a lattice. Following 
refs.\cite{Luck,Hermisson}, we discuss the initial sequences, 
the substitution rules, which generate the sequences, and the wandering 
exponents. Then we will show the stationary probability distributions of our 
stochastic model with each of these types of aperiodic disorder.
   
\subsubsection{Thue-Morse (TM) sequence}
The TM sequence $S=ABBABAAB \cdots$ is generated by the initial 
sequence $S_1=AB$ and the iterative substitution rules $A \rightarrow AB$ and 
$B \rightarrow BA$. Let $\#_n(A)$ and $\#_n(B)$ be the numbers of $A$ and $B$, 
respectively, in the sequence of the $n$-th generation $S_n$. From the 
substitution rules, $\#_{n+1}(A)$ and $\#_{n+1}(B)$ are obtained from
$\#_n(A)$ and $\#_n(B)$ as 
\begin{equation}
\left[
\begin{array}{c}
\#_{n+1}(A)\\
\#_{n+1}(B)
\end{array}
\right]
=M
\left[
\begin{array}{c}
\#_{n}(A)\\
\#_{n}(B)
\end{array}
\right],
\end{equation}
where the substitution matrix $M$ is given as 
\begin{equation}
M=
\left[
\begin{array}{cc}
1&1\\
1&1
\end{array}
\right].
\end{equation}
By diagonalizing the substitution matrix $M$, we find that the eigenvalues 
are $\lambda_1=2$ and $\lambda_2=0$, and 
\begin{eqnarray}
\#_{n+1}(A)+\#_{n+1}(B) &=& 2\{\#_{n}(A)+\#_{n}(B)\},
\label{TM1}
\\
\#_{n+1}(A)-\#_{n+1}(B) &=& 0.
\label{TM2}
\end{eqnarray} 
From Eqs.(\ref{TM1}) and (\ref{TM2}) we obtain 
\begin{equation}
\#_n(A) = \#_n(B) = 2^{n-1}.
\end{equation}
For the $n$-th generation sequence $S_n$ the length $L=\#_n(A)+\#_n(B)$ 
and the geometrical fluctuation $\Delta=\#_n(A)-\#_n(B)$ are $L=2^n$ and 
$\Delta=0$, respectively. From the definition of the wandering exponent 
$\Omega_{\rm TM}$, $\Delta \sim L^{\Omega_{\rm TM}}$ , it is obtained as the ratio 
of the logarithm of (the absolute value of) the second-largest eigenvalue 
to the logarithm of the largest eigenvalue,   
\begin{equation}
\Omega_{\rm TM} = \frac{\log|\lambda_2|}{\log \lambda_1} = -\infty.
\label{TMwander}
\end{equation} 
Since $\#_n(A)=\#_n(B)$ holds in any generation, 
the vanishing drift velocity condition is $b=a^{-1}$. 

\begin{figure}
\begin{center}
\includegraphics[width=8cm]{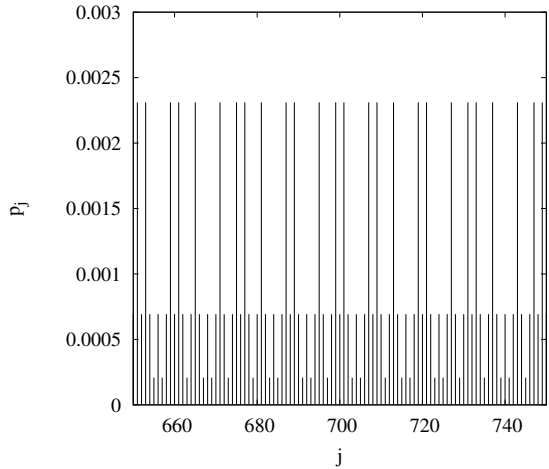}
\caption{Stationary probability distribution $\{p_j\}$ for the Thue-Morse 
model with $L=1024$ and $a=0.3$. For visibility, only the results of 
$650 \le j \le 750$ are shown. The distribution is extended. 
\label{tmdist}}
\end{center}
\end{figure}

Figure \ref{tmdist} shows the stationary probability distribution of the 
TM model with $L=1024$ and $a=0.3$. The diffusion of the TM model is known to be 
normal\cite{ITR}. We observe that the distribution is extended. 
From Eq.(\ref{solution}), the distribution is analytically obtained in a simple 
form: For the $n$-th sequence $S_n$ with a given $a$, the stationary probability 
distribution is composed of $2^{n-1}$ sites with $p_j=1/C$, 
$2^{n-2}$ sites with $p_j=a/C$ and $2^{n-2}$ sites with $p_j=1/aC$, 
where $C=2^{n-1}+(a+a^{-1})2^{n-2}$ is the normalization constant.
These three types of measures are aligned aperiodically. 

\subsubsection{Rudin-Shapiro (RS) sequence}
The RS sequence $S=AAABAABA \cdots$ is generated by the 
initial sequence $S_1=AA$ and the substitution rules $AA \rightarrow AAAB$, 
$AB \rightarrow AABA$, $BA \rightarrow BBAB$, and $BB \rightarrow BBBA$. 
In order to calculate $\#_n(A)$ and $\#_n(B)$ for the $n$-th sequence $S_n$, 
it is convenient to consider $\#_n(AA)$, $\#_n(AB)$, $\#_n(BA)$, and $\#_n(BB)$. 
Using them, we obtain
\begin{equation}
\left[
\begin{array}{c}
\#_{n+1}(AA)\\
\#_{n+1}(AB)\\
\#_{n+1}(BA)\\
\#_{n+1}(BB)
\end{array}
\right]
=M
\left[
\begin{array}{c}
\#_{n}(AA)\\
\#_{n}(AB)\\
\#_{n}(BA)\\
\#_{n}(BB)
\end{array}
\right],
\label{recursion}
\end{equation}
where the $4 \times 4$ substitution matrix is
\begin{equation}
M=
\left[
\begin{array}{cccc}
1&1&0&0\\
1&0&1&0\\
0&1&0&1\\
0&0&1&1
\end{array}
\right].
\end{equation}
By diagonalizing the substitution matrix $M$, we find that the eigenvalues 
are $2$, $\pm \sqrt{2}$, and $0$, and for the first three eigenvalues, 
\begin{eqnarray}
&&\#_{n+1}(AA) + \#_{n+1}(AB) + \#_{n+1}(BA) + \#_{n+1}(BB)
\nonumber
\\
&&= 2\{\#_n(AA) + \#_n(AB) + \#_n(BA) + \#_n(BB)\},
\label{RS1}
\\
&&(\sqrt{2}+1)\#_{n+1}(AA) + \#_{n+1}(AB) - \#_{n+1}(BA) - (\sqrt{2}+1)\#_{n+1}(BB)
\nonumber
\\
&&= \sqrt{2}
\{(\sqrt{2}+1)\#_{n}(AA) + \#_{n}(AB) - \#_{n}(BA) - (\sqrt{2}+1)\#_{n}(BB)\},
\label{RS2}
\\
&&(\sqrt{2}-1)\#_{n+1}(AA) - \#_{n+1}(AB) + \#_{n+1}(BA) - (\sqrt{2}-1)\#_{n+1}(BB)
\nonumber
\\
&&= -\sqrt{2}
\{(\sqrt{2}-1)\#_{n}(AA) - \#_{n}(AB) + \#_{n}(BA) - (\sqrt{2}-1)\#_{n}(BB)\}.
\label{RS3}
\end{eqnarray}
From Eqs.(\ref{RS1})-(\ref{RS3}),
\begin{eqnarray}
\#_{n+1}(A)+\#_{n+1}(B) &=& 2\{\#_{n}(A)+\#_{n}(B)\},
\label{RS4}
\\
\#_{n+1}(A)-\#_{n+1}(B) &=& 2^{\lceil n/2 \rceil}\{\#_{n}(A)-\#_{n}(B)\},
\end{eqnarray} 
where $\lceil n/2 \rceil$ is the ceiling function:
\begin{equation}
\lceil n/2 \rceil =
\begin{cases}
n/2, &n\text{ even}\\
(n+1)/2. &n \text{ odd} 
\end{cases}
\end{equation}
Then we immediately find that  
\begin{equation}
\begin{split}
\#_n(A)=2^{n-1}+2^{\lceil n/2 \rceil-1},
\\
\#_n(B)=2^{n-1}-2^{\lceil n/2 \rceil-1},
\end{split}
\end{equation}
and the length and geometrical fluctuation of the $n$-th sequence are 
$L=2^n$ and $\Delta=2^{\lceil n/2 \rceil}$, respectively. This last result means that 
the geometrical fluctuation scales as $\Delta \sim 2^{n/2}$, which 
corresponds to the second-largest eigenvalues, $\pm \sqrt{2}$.

The wandering exponent $\Omega_{\rm RS}$ is
\begin{equation}
\Omega_{\rm RS}=\frac{\log \sqrt{2}}{\log 2}=\frac{1}{2}.
\end{equation}
The geometrical fluctuation of the RS sequence grows unboundedly with the 
sequence length as $\Delta \sim L^{1/2}$.  Note that the value 
$\Omega_{\rm RS}=1/2$ coincides with that of the random binary sequence. 

\begin{figure}
\begin{center}
\includegraphics[width=8cm]{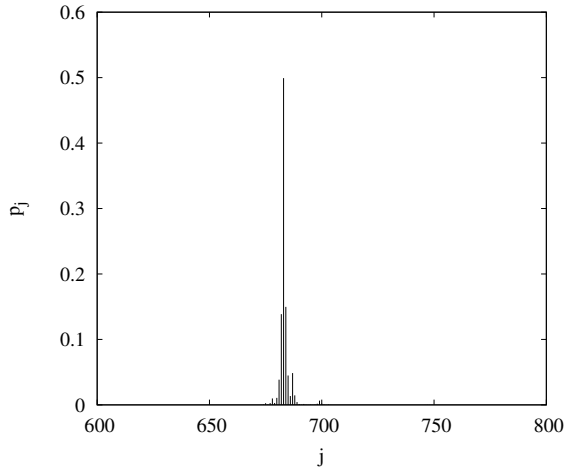}
\caption{Stationary probability distribution for the Rudin-Shapiro model 
with $L=1024$ and $a=0.3$. For visibility, Only the results of 
$600 \le j \le 800$ are shown. Clearly the distribution is localized.
\label{rsdist}}
\end{center}
\end{figure}

For the drift velocity to vanish, $b=a^{-\#_n(A)/\#_n(B)}$ for a given $a$. 
In the limit as $n \rightarrow \infty$, $b$ approaches $a^{-1}$. 
The stationary probability distribution of the RS model with $a=0.3$ and 
$L=1024$ is shown in FIG.\ref{rsdist}. The distribution is strongly localized. 
It was reported that diffusion is ultlaslow in the RS model, 
where the mean square displacement scales as Eq.(\ref{logdif})\cite{ITR}.  

\subsubsection{Paperfolding (PF) sequence}
The PF sequence $S=AABAABBA \cdots$ is generated by the initial 
sequence $S_1=AA$ and the substitution rules $AA \rightarrow AABA$, 
$AB \rightarrow AABB$, $BA \rightarrow ABBA$, and $BB \rightarrow ABBB$. 
The recursion relation for the numbers of letters $A$ and $B$ 
is expressed in the same form as Eq.(\ref{recursion}), where the substitution 
matrix in this case is given as
\begin{equation}
M=
\left[
\begin{array}{cccc}
1&1&0&0\\
0&0&1&1\\
1&0&1&0\\
0&1&0&1
\end{array}
\right].
\end{equation}
The eigenvalues of $M$ are $2$, $1$, and $0$ (which is doubly degenerate), 
and for the first two eigenvalues, 
\begin{eqnarray}
&&\#_{n+1}(AA) + \#_{n+1}(AB) + \#_{n+1}(BA) + \#_{n+1}(BB)
\nonumber
\\
&&= 2\{\#_n(AA) + \#_n(AB) + \#_n(BA) + \#_n(BB)\},
\label{PF1}
\\
&&\#_{n+1}(AA) - \#_{n+1}(AB) + \#_{n+1}(BA) - \#_{n+1}(BB)
\nonumber
\\
&&= \#_n(AA) - \#_n(AB) + \#_n(BA) - \#_n(BB),
\label{PF2}
\end{eqnarray}
We find that for the $n$-th sequence $L=2^n$, $\Delta=2$, and 
\begin{equation}
\begin{split}
\#_n(A)=2^{n-1}+1,
\\
\#_n(B)=2^{n-1}-1.
\end{split}
\end{equation}
The wanderling exponent of the PF sequence vanishies:
\begin{equation}
\Omega_{\rm PF} = \frac{\log 1}{\log 2}=0.
\end{equation}
In fact the geometrical fluctuation grows logarithmically with $L$ although 
it remains constant at the endpoint. 

\begin{figure}
\begin{center}
\includegraphics[width=8cm]{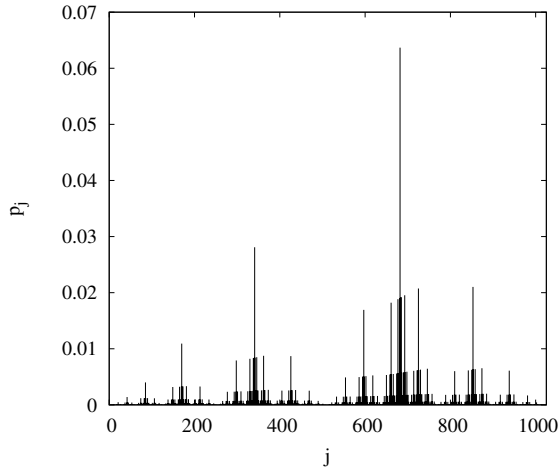}
\caption{Stationary probability distribution for the paperfolding model 
with $L=1024$ and $a=0.3$. A complex and hierarchical structure ca be observed.
\label{pfdist}}
\end{center}
\end{figure}

For the drift velocity to vanish, 
\begin{equation}
b=a^{-(2^{n-1}+1)/(2^{n-1}-1)}, 
\end{equation}
for a given $a$, which converges to $a^{-1}$ in the limit as 
$n \rightarrow \infty$. It was found that in the PF model, the 
diffusion is anomalous, and it is written as in Eq.(\ref{anomdif})\cite{ITR}, 
where the exponent $\phi$ depends on the inhomogeneity parameter $a$. 
The stationary probability distribution of the PF model with $a=0.3$ and 
$L=1024$ is shown in FIG.\ref{pfdist}. The distribution appears to be singular, 
and appears to be neither extended nor localized. We can observe its complex and 
hierarchical structure. 

\section{Multifractal analysis}

\subsection{Formulation: on a one-dimensional support}
\label{formulation}
Aperiodic chains are defined in the limit as $L \rightarrow \infty$. 
Thus we estimate the results of the system as $L \rightarrow \infty$ 
by systematically extrapolating from the results of systems with finite $L$. 

Let us review the formulation of multifractal on a one-dimensional 
support\cite{HJKPS,HK}.
Suppose that a stationary probability measure for a finite one-dimensional 
$L$-site system $\{p_j\}_{j=1,2,\cdots,L}$ is given. 
The partition function $Z(q,L)$ is introduced as
\begin{equation}
Z(q,L)=\sum_{j,p_j \ne 0} (p_j)^q.
\label{partition}
\end{equation}
The multifractal exponent for the finite system $\tau(q,L)$ is defined as
\begin{equation}
\tau(q,L)=-\frac{\log Z(q,L)}{\log L}.
\label{tau_finite}
\end{equation}
By the Legendre transformation, the singularity exponent $\alpha$ for the 
finite system and its fractal dimension $f(\alpha)$ are obtained, as functions 
of $q$ and $L$:
\begin{eqnarray}
\alpha(q,L) &=& \frac{\partial \tau(q,L)}{\partial q},
\label{alpha_finite}
\\
f(\alpha(q,L)) &=& q\alpha(q,L)-\tau(q,L).
\label{f_finite}
\end{eqnarray}  
However it is not practical to evaluate $\alpha$ and $f(\alpha)$ numerically 
from Eqs.(\ref{alpha_finite}) and (\ref{f_finite}), since this requires 
numerical differentiation, which may produce relatively large errors. 
Therefore it is better to evaluate them directly. We show this below,  
following the method presented in ref.\cite{CJ}.   

Let us construct a new probability measure $\{\mu_j(q)\}$ from $\{p_j\}$:
\begin{equation}
\mu_j(q) = \frac{(p_j)^q}{\sum_{j=1}^L (p_j)^q}.
\label{mu_construction}
\end{equation}
Then let us define $\zeta(q,L)$ and $\xi(q,L)$ as
\begin{eqnarray}
\zeta(q,L) &=& \sum^L_{j=1} \mu_j(q) \log p_j,
\label{zeta_finite}
\\
\xi(q,L) &=& \sum^L_{j=1} \mu_j(q) \log \mu_j(q),
\label{xi_finite}
\end{eqnarray}
from which we obtain $\alpha(q,L)$ and $f(\alpha(q,L))$ as 
\begin{eqnarray}
\alpha(q,L) &=& -\frac{\zeta(q,L)}{\log L},
\label{alphadef}
\\
f(\alpha(q,L)) &=& -\frac{\xi(q,L)}{\log L}.
\label{fdef}
\end{eqnarray}
Direct calculation shows that the definitions Eqs.(\ref{alphadef}) and 
(\ref{fdef}) satisfy the relations Eqs.(\ref{alpha_finite}) and 
(\ref{f_finite}). Note that the above formulation has some similarities 
with the thermodynamic formulation of the R{\'e}nyi entropy $H(q)$ of 
dynamical systems, which is defined as\cite{RenyiEntropy}
\begin{equation}
H(q)=\frac{1}{1-q}\log \left[\sum_j(p_j)^q\right],
\label{Renyientropy}
\end{equation} 
for $q>0$.
The phase transition related to the R{\'e}nyi entropy of a 
deterministic chaotic system is discussed in ref.\cite{RV}. 

Next we estimate the finite size effect. For example, the "true" 
value of $\tau(q)$ for a system of infinite size is defined as 
\begin{equation}
\tau(q) = \lim_{L \rightarrow \infty} \tau(q,L),
\label{limitdef}
\end{equation}  
and $\alpha(q)$ and $f(\alpha(q))$ are defined similarly. 
They should be obtained by careful 
extrapolation from the results for systems of finite size. 
From Eqs.(\ref{partition}), (\ref{tau_finite}) and (\ref{limitdef}), 
we expect that 
\begin{equation}
\tau(q)-\tau(q,L)=\mathcal{O}(1/\log L).
\label{logcorrection}
\end{equation}
Therefore we estimate the value of $\tau(q)$ from the plot of $\tau(q,L)$ 
against $1/\log L$ and the extrapolation to $1/\log L \rightarrow 0$. 

The localization property of a given distribution can be read from its 
multifractal spectrum, especially the results for $q \rightarrow \pm \infty$. 
This is known in quantum localization problem, where the probability 
distribution is given as the squared norm of the wavefunction\cite{HK}.
Let $\alpha_{\rm min}$ and  $\alpha_{\rm max}$ 
be $\alpha(q \rightarrow \infty)$ and $\alpha(q \rightarrow -\infty)$,
respectively, and let $f_{\rm min}=f(\alpha_{\rm min})$ and 
$f_{\rm max}=f(\alpha_{\rm max})$. 
For an extended distribution, the multifractal 
spectrum of the systems of finite size converges to a single point 
$\alpha = f = 1$ in the limit as $L \rightarrow \infty$. 
For a localized distribution, $\alpha_{\rm min}$ and $f_{\rm min}$ converge to $0$,
$\alpha_{\rm max}$ diverges to infinity, and $f_{\rm max}$ converges to unity. 
For a singular distribution, $\alpha_{\rm min}$ and $\alpha_{\rm max}$ take 
different finite values. The spectrum $f(\alpha)$ is a continuous and convex 
curve which takes values only within 
$\alpha \in [\alpha_{\rm min},\alpha_{\rm max}]$.
 
\subsection{Localization and inverse participation ratio}
\label{secipr}
Note that the partition function Eq.(\ref{partition}) with $q=2$, 
$Z(q=2,L)$ is equivalent to  the "inverse participation ratio (IPR)" . 
The IPR was originally introduced in quantum localization 
problem\cite{BD,Thouless} more than forty years ago. 
Its purpose is to simply evaluate the localization property of a given 
distribution. It has been used not only 
in quantum mechanics but also even in finance\cite{PGRNAS}. 
And some generalizations have been attempted recently\cite{MWA,YYL}.    
The scaling behavior of the IPR against the system size $L$ is used to classify 
the localization property of a given distribution. 
If a state is extended, the IPR is inversely proportional to 
the system size, since the probability measure at a site is roughly inversely 
proportional to the system size, {\it i.e.}, $p_j \sim L^{-1}$. 
On the other hand, if a state is localized, the IPR is almost independent of 
the system size.  If a state is singular, which is called "critical" 
in the context of quantum localization, the scaling behavior of 
the IPR is intermediate between the behavior in the above two cases:   
\begin{equation}
IPR(L) \sim L^{-\delta} \quad \text{with }0<\delta<1.
\label{iprscaling}
\end{equation}
We expect that these scaling behaviors hold for the stationary probability 
distribution of our classical stochastic models.

\begin{figure}
\begin{center}
\includegraphics[width=8.5cm]{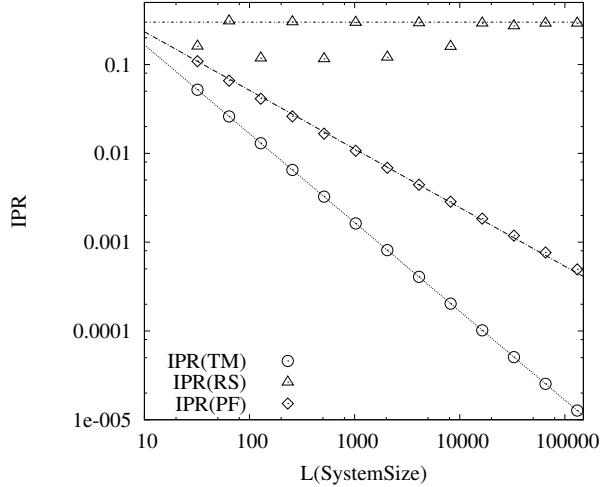}
\caption{Log-log plots of the inverse participation ratio $Z(2,L)$ against 
system size $L$ for the Thue-Morse, Rudin-Shapiro, and paperfolding models 
with $a=0.3$. Lines are, from top to bottom, $Const.$, $L^{-0.69}$, 
and $L^{-1}$.  
\label{ipr}}
\end{center}
\end{figure}

\begin{figure}
\begin{center}
\includegraphics[width=8cm]{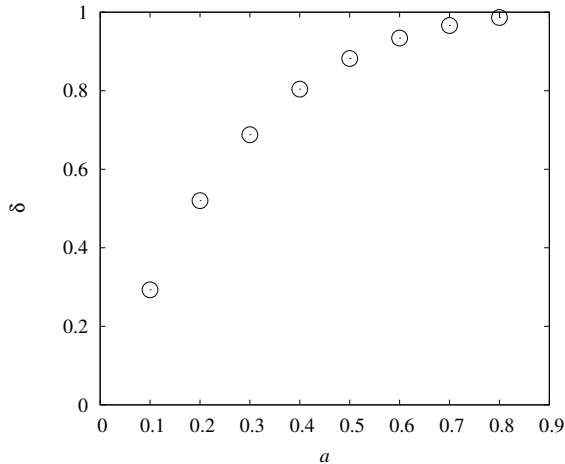}
\caption{Plots of the exponent $\delta$ in Eq.(\ref{iprscaling}) against $a$. 
In the homogeneous limit $a \rightarrow 1$, $\delta \rightarrow 1$. 
\label{ppfipr}}
\end{center}
\end{figure}

Figure \ref{ipr} shows the log-log plots of the IPR against the system size 
$L$ for the TM, RS, and PF models with $a=0.3$. 
For the TM model, it is observed that $Z(2,L) \sim L^{-1}$. 
For the RS model, for small $n$, the IPR depends on whether $n=\log_2L$ is 
even or odd. However, as $n$ increases,  the series of the results for odd $n$ 
converge with those for even $n$, and the results become independent of the 
system size, {\it i.e.}, $Z(2,L) \sim Const.$ 
These results are consistent with the localization properties of their 
probability distributions as shown in FIGs.\ref{tmdist} and \ref{rsdist}, 
which are extended for the TM model and localized for the RS model, respectively. 
These scaling properties are independent of the value of $a$.    
For the PF model with a singular probability distribution,  
the scaling behavior is $Z(2,L) \sim L^{-\delta}$ with $\delta=0.69$. 
Note that in this case, the exponent $\delta$ depends on $a$. 
Figure \ref{ppfipr} shows the 
$a$-dependence of $\delta$. It monotonically increases with $a$ and 
approaches $1$ as $a \rightarrow 1$, 
since the system with $a=1$ is homogeneous and so the distribution is 
extended.  
 
\subsection{Multifractal Spectra}
Since $\alpha_{\rm max}$ and $f_{\rm max}$ are dominated by the 
smallest measure of the distribution, they may have large numerical errors.
Therefore we will restrict our discussion to only $\alpha_{\rm min}$ and 
$f_{\rm min}$.
 
\begin{figure}
\begin{center}
\includegraphics[width=8cm,angle=0]{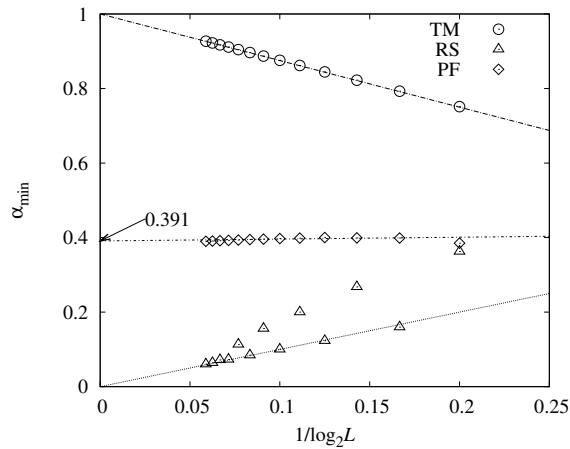}
\caption{Plots of $\alpha_{\rm min}(L)$ against $1/n = 1/\log_2 L$ for the 
Thue-Morse, Rudin-Shapiro, and paperfolding model with $a=0.3$. 
Their linear dependence means that the leading 
correction is $\mathcal{O}(1/\log L)$. 
\label{amin}}
\end{center}
\end{figure}

\begin{figure}
\begin{center}
\includegraphics[width=8cm]{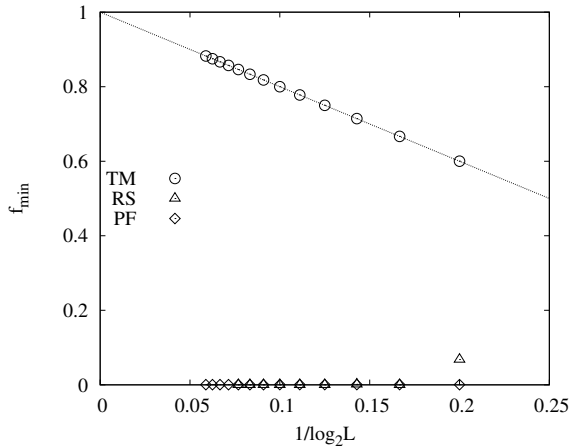}
\caption{Plots of $f_{\rm min}(L)=f(\alpha_{\rm min}(L))$ against $1/\log_2 L$ 
for the Thue-Morse, Rudin-Shapiro, and paperfolding model with $a=0.3$. 
The leading correction here is also $\mathcal{O}(1/\log L)$. 
\label{fmin}}
\end{center}
\end{figure}

Figure \ref{amin} shows the plots of $\alpha_{\rm min}(L)$ against 
$1/n = 1/\log_2 L$ for the TM, RS, and PF models with $a=0.3$. 
For the TM and PF models, these plots are linear. For the RS model, 
similar to the case of the scaling of the IPR, some parity dependence is 
found for small $n$. However, for large $n$ the plots become linear, 
independent of the parity of $n$. This validates our expectation of the 
finite-size effect, Eq.(\ref{logcorrection}).  
Extrapolating the plots toward $1/\log L \rightarrow 0$, we find that 
$\alpha_{\min} \rightarrow 1$ for the TM model, $\alpha_{\min} \rightarrow 0$ 
for the RS model, and $\alpha_{\min} \rightarrow \sim 0.391$ for the PF model.  
Figure \ref{fmin} shows the plots of $f_{\rm min}(L)$ against $1/\log_2 L$. 
It can be observed that in the limit as $1/\log L \rightarrow 0$, 
$f_{\min} \rightarrow 1$ for the TM model, and $f_{\min} \rightarrow 0$ 
for the RS and PF models.    
These results are consistent with the criteria mentioned in the last 
paragraph of Sec.\ref{formulation} and also with the 
results of the IPR scaling behavior obtained in Sec.\ref{secipr}.

\begin{figure}
\begin{center}
\includegraphics[width=8cm]{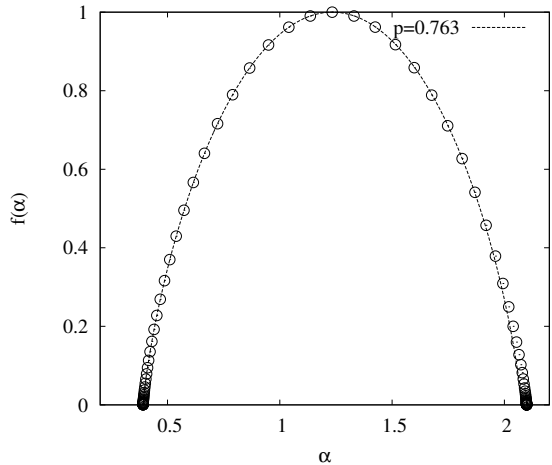}
\caption{Multifractal $f(\alpha)$ spectrum of the stationary probability 
distribution of the PF model with $a=0.3$ and that of the binomial branching 
process with $p=0.763$. 
\label{mf03}}
\end{center}
\end{figure}

For the PF model, the multifractal $f(\alpha)$ spectrum takes continuous 
values within $\alpha \in [\alpha_{\rm min},\alpha_{\rm max}]$, where 
$\alpha_{\rm min}$ and $\alpha_{\rm max}$ are both positive finite values. 
The multifractal spectrum for the PF model with $a=0.3$ is shown in 
FIG.\ref{mf03}. It is convex upwards, which is a universal property, and 
takes the maximum value $f=1$, reflecting the fact that the support of the 
probability distribution is one-dimensional. Moreover it looks symmetric with 
respect to $\alpha=\alpha_0$, where it takes the maximum.  
This result is in quite good agreement 
with the spectrum of the "binomial branching process", which is a 
simple process constructed by the recursion of elementary uneven 
partitioning. 
The binomial partitioning process was first introduced as a simple model for 
the hierarchical energy cascade of turbulence\cite{CMKRS}. It has been 
applied as a simple model for various systems including 
the sidebranch structure of a dendrite\cite{MH} and fragmentation\cite{KSH}. 
In this process, the spectrum can be calculated exactly due to its simplicity:
\begin{eqnarray}
\alpha(q)
= -\frac{\eta \log p + (1-\eta)\log(1-p)}{\log 2},
\label{bbpalpha}
\\
f(\alpha(q)) = -\frac{\eta \log \eta + (1-\eta)\log(1-\eta)}{\log 2},
\label{bbpf}
\end{eqnarray}
where
\begin{equation}
\eta = \frac{p^q}{p^q+(1-p)^q},
\label{eta}
\end{equation}
and $1/2<p<1$ is the partitioning parameter, which is the only free parameter 
in the process. From Eq.(\ref{bbpalpha}) we immediately obtain   
$\alpha_{\rm min}=-\log_2p$ and $\alpha_{\rm max}=-\log_2(1-p)$.
This agreement shows that, in the PF model, there exists a mechanism 
which partitions the probability measure unevenly and hierarchically, 
in a way similar to that in the binary branching process.
This is attributed to the fact that the effect of the fluctuation of 
the PF sequence, due to its vanishing wandering exponent, is almost 
independent of length scale. 

\begin{figure}
\begin{center}
\includegraphics[width=8cm]{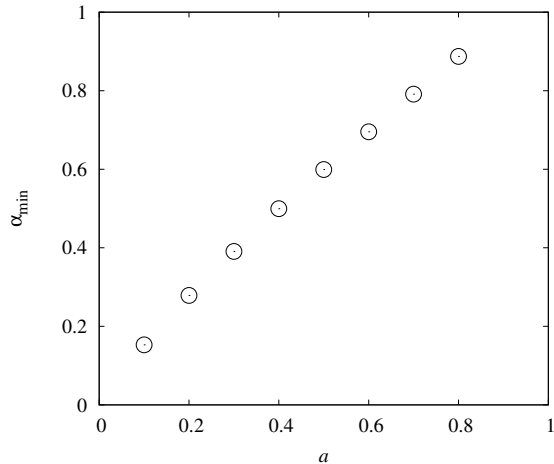}
\caption{Plot of $\alpha_{\rm min}$ against $a$. The value of 
$\alpha_{\rm min}$ converges to unity 
in the homogeneous limit as $a \rightarrow 1$.
\label{aminppf}}
\end{center}
\end{figure}

Figure \ref{aminppf} shows the $a$-dependence of $\alpha_{\rm min}$. 
We find that $\alpha_{\rm min}$ is  a monotonically increasing function of $a$. 
In the limit as $a \rightarrow 1$, the system becomes homogeneous, and 
therefore the distribution is extended and $\alpha_{\rm min}$ approaches 
unity, which characterizes an extended distribution. 
Since the multifractal spectra for the PF model and the binary 
partitioning process are in good agreement, the parameter in the binary 
branching process, $p$, and $a$ are related as 
$\alpha_{\rm min}(a)=-\log_2p$.  

Thus far, we have restricted ourselves to the case of $a<1$ for each model. 
For $a>1$, at least the localization property of the probability 
distribution and the multifractal spectra do not vary under 
$a \leftrightarrow a^{-1}$, in the limit as $L \rightarrow \infty$. 
This is probably a consequence of the fact that in the underlying aperiodic 
sequence, the ratio of the number of $A$ to that of $B$ converges to unity. 

\section{Summary and outlook}
We found that the stationary probability distribution of a random walk on 
a one-dimensional aperiodically disordered lattice shows a characteristic 
localization pattern which corresponds to its diffusional behavior. 
The results are summarized in TABLE \ref{table_summary}.
The localization pattern of the distribution (extended, localized, or 
singular) depends on the wandering exponent of the background aperiodic 
sequence. These types of pattern can be distinguished by the finite-size  
scaling of the partition function $Z(q=2,L)$, the singular exponent 
$\alpha_{\rm min}$, and the fractal dimension $f_{\rm min}$. 
In particular, for the distribution of the model with a vanishing wandering 
exponent, we obtained a continuous multifractal spectrum with finite 
$\alpha_{\rm min}$ and $\alpha_{\rm max}$ ($\alpha_{\rm min} \ne \alpha_{\rm max})$.
This spectrum reflects the singular and hierarchical structure of the 
distribution and agrees well with the spectrum of the binomial branching 
process. 

\begin{table}
\begin{tabular}{l|ccccc}
Sequence&$\Omega$&PDF&Diffusion&$\alpha_{\rm min}$&$f_{\rm min}$\\
\hline
TM&-&extended&normal&1&1\\
RS&+&localized&ultraslow&0&0\\
PF&0&singular&anomalous&finite&0\\
\end{tabular}
\caption{Summary of the results. $\Omega$ denotes the wanderling exponent, 
PDF denotes the stationary probability distribution function, and 
$\alpha_{\rm min}$ and $f_{\rm min}$ denote 
$\alpha(q\rightarrow \infty,L\rightarrow \infty)$ and 
$f(q \rightarrow \infty, L\rightarrow \infty)$, respectively. 
\label{table_summary}}
\end{table}

We considered only the case with a vanishing drift velocity $v_d=0$, since we 
were interested in the diffusional behavior. As mentioned in Section \ref{1drw}, 
a finite drift velocity $v_d \ne 0$ causes a finite current through the lattice 
and makes the distribution extended. It may be an interesting problem to 
determine how a localized or singular distribution changes by a finite 
drift velocity.       

\begin{acknowledgments}
The author would like to thank Professor H.Honjo for useful comments.
\end{acknowledgments}
   
\appendix*
\section{Binomial branching process}
We discuss the binomial branching process\cite{CMKRS} so that this article is 
self-contained. Its multifractal spectrum can be exactly calculated due to its 
simple structure.
\begin{figure}
\begin{center}
\includegraphics[width=8cm]{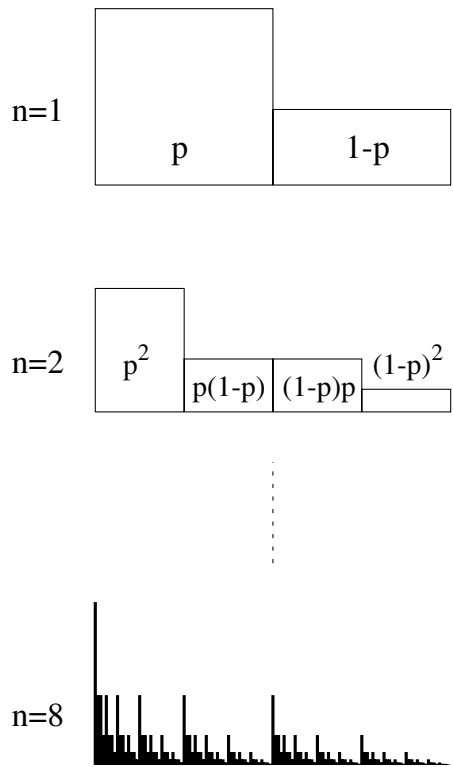}
\caption{Different stages of the binomial branching process. Each segment is 
divided into two equal subsegments at the next stage and its measure is 
divided into nonequal fractions, $p$ and $(1-p)$. This figure is cited from
ref.\cite{MH}. \label{bpp}}
\end{center}
\end{figure}

Suppose that a segment of length $1$ is divided into two segments of length 
$1/2$. A probability measure $p>1/2$ is assigned to the left segment and 
$(1-p)$ to the right. This $p$ is the only free parameter of the process. 
Next, each segment is subdivided into two equal halves and the measure is 
partitioned into $p$ to the left and $(1-p)$ to the right. There are now four 
segments, each of length $1/4$, and the measures $p^2$, $p(1-p)$, $(1-p)p$, 
and $(1-p)^2$ are assigned to the segments from left to right. 
This procedure is iterated (see FIG.\ref{bpp}), which shows the hierarchical 
structure $n=8$. At the $n$-th stage, there are $2^n$ segments, each of 
length $2^{-n}$ and the number of segments with measure $p^k(1-p)^{n-k}$, 
$k=0,1,\cdots ,n$, is $\binom nk = n!/[k!(n-k)!]$. Therefore the partition 
function for this stage, $Z(q,n)$ is immediately obtained as 
\begin{eqnarray}
Z(q,n) &=& \sum_{k=0}^n \left(
\begin{array}{c}
n
\\
k
\end{array}
\right)
[p^k(1-p)^{n-k}]^q
\nonumber
\\
&=&[p^q+(1-p)^q]^n.
\end{eqnarray}
The multifractal exponent $\tau(q)$ is, in the limit as $n \rightarrow \infty$:
\begin{equation}
\tau(q) = -\frac{\log[p^q+(1-p)^q]}{\log 2}.
\end{equation}
From this and by using the Legendre transformation the singularity exponent 
$\alpha(q)$ and the fractal dimension $f(\alpha(q))$ are obtained as 
Eqs.(\ref{bbpalpha}) and (\ref{bbpf}). The direct evaluation, 
Eqs.(\ref{alphadef}) and (\ref{fdef}), gives the same result.     
The spectrum is symmetric with respect to 
$\alpha_0=-[\log_2p+\log_2(1-p)]/2$ and takes the maximum $f(\alpha_0)=1$, 
which reflect the fact that the support is one-dimensional.

\end{document}